\documentclass[sigconf]{acmart}

\usepackage{booktabs}
\usepackage[title]{appendix}
\usepackage{url}
\usepackage{pdfpages}
\usepackage[section]{placeins}
\usepackage{subcaption}
\usepackage{dirtytalk}
\renewcommand\footnotetextcopyrightpermission[1]{}

\begin{document}
\pagestyle{plain}
\setcounter{page}{0}

\includepdf{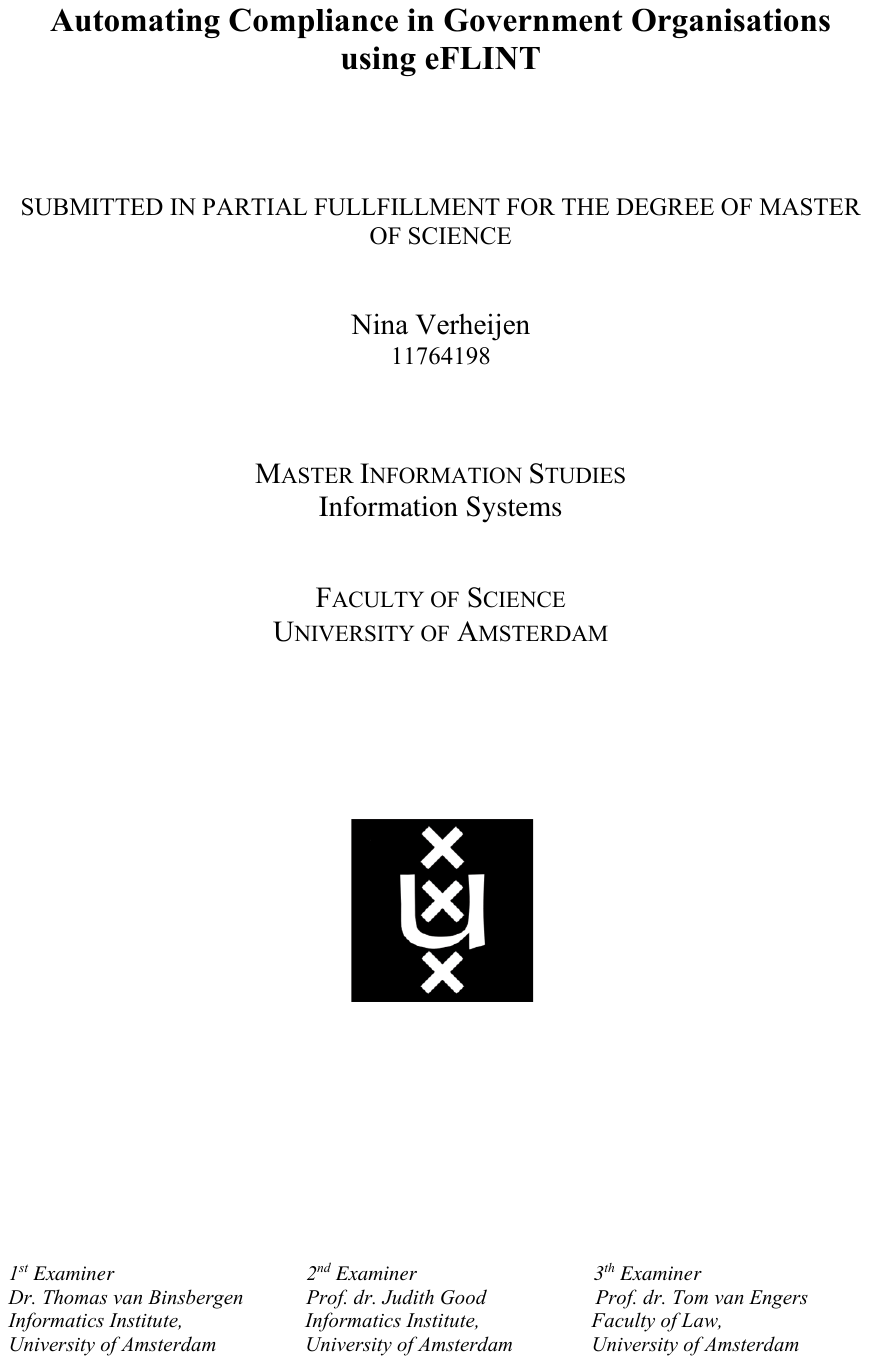}

\title{Automating Compliance in Government Organisations using eFLINT}

\author{Nina Verheijen}
\affiliation{
  \institution{University of Amsterdam}
  \city{Amsterdam} 
  \country{The Netherlands}
  }
\email{Nina.Verheijen@student.uva.nl}

\begin{abstract}

Ensuring compliance of norms and policies when working on administrative law cases can be difficult to manage for government organisations. Automating this process could save a lot of time, effort and ensure compliance. Prior research resulted in a method to formalize sources of norms. These can be turned into executable specifications using the domain-specific language eFLINT, which can be used for automating compliance. However, the current interface of eFLINT prevents adaption by legal experts. 

The aim of this research was to bridge this gap by developing a prototype based on eFLINT, for automating compliance within government organisations. To get a better understanding of the needs and requirements of potential users, qualitative research was conducted. This consisted of semi-structured interviews to gather requirements, which were analyzed using a thematic analysis method. Based on the analyzed data, a design for the interface of the prototype was made. The final prototype was evaluated in a user end study which included a cognitive walkthrough and user testing. The prototype proved to be a good first step in the right direction with a lot of room for further development.
\end{abstract}


\maketitle

\section{Introduction} \label{Introduction}


Government organisations cope with many norms and policies when dealing with administrative law cases. Making sure these policies are being enforced is for the most part done manually. This can get complicated and time consuming with ever-changing laws and varying cases. Manually monitoring and enforcing all the different policies can also be prone to errors. By automating this process these complications can be prevented.

Prior research resulted in a method to make formal interpretations of normative sources. These formal interpretations can be turned into executable specifications using executable FLINT (eFLINT), a domain-specific language (DSL)~\cite{VanBinsbergen2020EFLINT:Specifications}. eFLINT stands out to other formalizing languages in two ways. It enables formalization on different levels of legal documents, from smart contracts to events triggered in software systems. The language can also be used to keep track of changes in a case as it develops over time due to the actions performed by involved parties. In addition, government organisations use software to keep track of a case as it develops over time and automate certain actions. eFLINT can be used to help automate compliance of norms and policies in these software systems.

Although eFLINT could potentially be useful for government organisations, it's current interface prevents adoption by legal experts~\cite{L.T.vanBinsbergenandG.Sileno2020Web-interfaceEFLINT.}. Therefore, the aim is to better understand the needs of potential users, in order to design an interface that meets their needs. This leads to the following research question:\\



\textit{What interface is required for a system based on eFLINT, that is used for automating compliance of norms and policies within government organisations?} \\

To be able to answer this research question, the following sub-questions were formulated: \\

1. Which types of use do potential users encounter and are there similar patterns between different organisations? \\
\indent 2. What needs and requirements do potential users have for the prototype? \\

In order to establish the potential of eFLINT, a prototype was developed to demonstrate government organisations how eFLINT could be of use to them. The prototype was not designed for a specific organisation, but should be generally applicable. Before implementing such prototype, the needs and requirements of potential users had to be gathered. This was done by conducting semi-structured interviews, in which questions were asked to get a clear picture of the use cases of the users and what information they need to work on their cases. An example of a use case could be a decision officer that receives requests for a subsidy. Here, eFLINT could support the decision officer by presenting what actions needs to be taken, based on applicable norms and information provided by the client.

Based on the requirements and use cases that came from the interviews, a prototype was developed. The prototype was tested by users to establish its user friendliness and make sure it is fit for purpose.

\section{Related Work} \label{Related Work}

To be able to design an interface for the prototype, a basic understanding of eFLINT was necessary to know its features and limitations. This section  covers the most important aspects of eFLINT.

    \subsection{About eFLINT} \label{eFLINT}
    
    
    Services provided by government organisations and companies are primarily defined in laws and regulations, but there is often a gap between the software systems that support these services and the laws and regulations that are applicable. This gap makes it difficult to keep up with changes in regulations and policies.
    
    eFLINT is a DSL that can be used for formalizing norms as executable specifications and is based on the legal framework of Hohfeld~\cite{VanBinsbergen2020EFLINT:Specifications}\cite{Hohfeld1913SomeReasoning}. An important aspect of eFLINT is that the language is action-based and supports the legal concept of power, meaning it has the ability to give or remove permissions or duties from actors. The benefit of an action-based approach is that  both software implementations and scenarios are action-based, which makes it easier to check the compliance. Normative relations can change over time by the effects of actions and events. eFLINT can simulate real-world situations and can be used to reason about the compliance of running administrative law cases. With eFLINT it is possible to track these changes, which means it is possible to see if and where exactly a violation occurs. The language supports manual exploration and enables external systems to trigger specific actions and events. It can be broadly applied since it also allows the redefinition of types to form specialized domains, which means that it is possible to reuse a norm specification across different applications in which certain terms might have a different meaning. All these features make it suitable for automating compliance of norms and policies within government organisations.


\section{Methodology} \label{Methodology}

Looking at the methodology of this research, it becomes clear that the project can be divided into two phases: the first phase being the process of conducting interviews and requirement elicitation, the second face being the designing, implementation and testing of the prototype. This section describes the methodology of both phases. The experimental setup and results of both phases are each described in order in sections \ref{Experimental Setup} - \ref{Results: Testing the prototype}.

eFLINT could be used to bridge the gap between law and computer, jurist and programmer, but to do so it needs to be accessible to the domain experts. Including users in early stages of development prevents incorrect assumptions being used whilst developing the prototype and allows requirement elicitation~\cite{Martin2012APerspective}. Therefore, designing the interface was done using a user-centred design (UCD) method, meaning that the users were involved in the designing process. The first step of UCD is to understand the context in which the users would use the system, more specifically, get a better understanding of the different use cases and the steps involved. 

Interviews were conducted to gather the needs and requirements for the interface. In a study where a similar methodology was used, a brainstorm session was done before the interviews to identify the research objectives that they wanted to get from the interviews~\cite{Martin2012APerspective}. In this research, a similar brainstorm session was held before the interviews. The general pattern of a use case was written down to establish which themes needed to be covered during the interviews. The interviews were semi-structured. This way the information that was needed to answer the questions from the themes described in section \ref{pattern} was acquired, while also leaving room for the participants to add additional insights gathered from the interview questions~\cite{DiCicco-Bloom2006TheInterview}. Since semi-structured interviews can be used as the sole data source, additional observations were not necessary~\cite{Adams2002PrimaryHard}. To select candidates for the interviews, a purposeful sampling method was used~\cite{Koerber2008QualitativeCommunicators}.

The results from the interview were analyzed, with the pattern described in section \ref{pattern} as guidelines. The pattern already contained the themes that were needed to categorize the data from the interview. Finding themes in data can be done using different qualitative methods. Some methods, however, are fixed to a particular position and have limited variability in how the method can be applied within that framework, such as conversation analysis (CA)~\cite{Hutchby1998ConversationApplications} or grounded theory~\cite{Strauss1998BasicsTechniques}. However, for this research a method was required that was not bound to a theory and that can be applied across a range of theoretical approaches. Such a method is thematic analysis~\cite{Braun2006UsingPsychology}. Thematic analysis has a flexible framework that can be applied in many different scenarios, but still gives enough guidance to ensure correct usage.

Based on the analyzed data from the interview, an interface was designed. The design was used to implement the prototype, which was made with the purpose of demonstrating government organisations how eFLINT could be of use to them. The prototype was not for one specific organisation, but served a general purpose. 

The prototype was evaluated by conducting an expert evaluation followed by usability testing along with interview questions. The expert evaluation was done by performing a cognitive walkthrough to find possible issues prior to the usability testing~\cite{Rieman1995UsabilityWalkthrough}. The usability testing ensured that the interface was user friendly, intuitive and gave insight in whether the prototype was fit for purpose. Lastly, the interview questions were used to reflect on the usability test and gave room for the users to give feedback.

\section{Experimental Setup: Establishing requirements} \label{Experimental Setup}

In this section the approach mentioned in section \ref{Methodology}, with regard to establishing the requirements, is explained in more detail. Topics that are covered in this section include: describing the pattern that occurs in use cases, the participants, the interviews and the analysis of the interview results.

    \subsection{Describing the pattern of a use case} \label{pattern}
    
    To create structure in the information that was gathered from the interviews, a pattern was created that covered the different themes that occur in a use case. These themes were: \textit{'Involved people'}, \textit{'Types of cases'}, \textit{'Gathering information'}, \textit{'Actions and duties'} and \textit{'Ending of a case'}. This section describes each of the themes in more detail.
    
    \subsubsection{Involved people} \label{Involved people}
    When starting a case, the first step is to determine who the involved parties are, both on the client side and on the side of the organisation that takes on the case. For the company side, the question is how many people are working on a case. If there are multiple people working on the same case it should be clear who has which tasks and responsibilities. On the client side, the question is whether the client is a company or a civilian, and when it is a company, whether there is a specific person that represents the company. There can also be a conflict within the organisation about the definition of a norm, or a client disagrees with the decision that has been made. Another client can also get involved with a case later on. Will this continue as one complex case or will it be split into two cases that are still connected?
    
    \subsubsection{Types of cases} \label{Type of case}
    The next step is to determine the type of case, which might be dependent on the type of client. In a case where the client is a company instead of a civilian, different laws or policies might apply. The process could contain extra steps and more information might be needed. The type of case could also depend on what kind of request is being made. The process for a request for subsidy will be different than the process for an objection. For this theme the goal is to look for certain patterns that either occur in most cases, or that distinguish them. Finding those patterns can help to automate parts of a use case.
    
    \subsubsection{Gathering information} \label{Information gathering}
    Next, the required information should be gathered. What information is needed might depend on the type of case. The question here is how the information will be acquired in order to be able to analyse the information in the next stage. Perhaps there is a set list of statements that the decision officer uses to decide about a case? Will this be under supervision of the case manager or can the client do this independently? Is all information that is possibly needed gathered at once, or first only the necessary information and more information is added as needed later on?
    
    \subsubsection{Actions and duties} \label{Actions}
    When the necessary information is acquired, there will be a list of actions or duties that need to be completed, either by the client or by the case manager. A case can change over time, so new information could be needed, or a form would need to be filled in. If there are duties, there will also be a condition that determines whether that duty is fulfilled or violated. If the duty is violated, there might be consequences, which means there should also be someone to act upon those consequences.
    
    Actions can have three different states: allowed, not allowed, or unknown. The latter meaning that there is not enough information to know whether the action is allowed or not, or there might be a conflict between a law and the company policy for example. Sometimes an employee might want to differ from the normal path, because of special circumstances, and perform an action that is not allowed. In this case the question is who from the involved people has the authority to do this and if a motivation should be given. 
    
    Each action or duty usually comes from a legal source. This leads to the next point: to what extend can, or should, the source of an action be known to the client? Meaning, what is the level of transparency? This partly depends on how clear the company policy is, if the policy is complicated it might only confuse the client more. The level of transparency also depends on whether the client has access to information regarding the case.
    
    \subsubsection{Ending a case} \label{Case ending}
    
    The final theme is how a case ends, or whether it ends at all. This also raises questions about the data from the case, is this stored, and if yes, for how long.

    \subsection{The participants} \label{participants}
    
    The participants were potential users. This included people that were either involved with handling cases from civilians or organisations where laws and policies are involved, or people who were involved with the process of turning laws and policies into actions, which could also mean actions in software systems. The prototype should be generally applicable, thus the participants should come from different organisations in order to get different perspectives. This way the needs and requirements can be generalized and the interface will be usable for a wide range of government organisations. There were seven participants in total, from four different organisations. The participants that came from the same organisation had different functions. This should be kept in mind when analyzing the interviews to not create any bias towards one organisation. From the participants, two were not familiar with eFLINT, the other 5 participants were at least somewhat familiar eFLINT. 
    
    It should be noted that both during the interviews and user testing, ethical guidance was sought by the supervisors from this research. The interview protocol included an introduction asking for explicit consent to participate, as well as statements on ethical issues such as the data anonymisation and secure storage of data. The participants had the option to withdraw participation at any given moment and they did not have to answer questions they could not, or did not, want to answer.
    
    \subsection{First interview session} \label{First interview}
    
    The first set of interviews was semi-structured, this way the information that was needed for designing the interface was gathered, but it also left room to get new insights from the participants~\cite{Sharp2007InteractionInteraction}~\cite{DiCicco-Bloom2006TheInterview}. The interviews were conducted online. All participants were informed about eFLINT and the purpose of the prototype beforehand. As a preparation for the interview, the participants were asked to come up with a possible use case of eFLINT or the prototype within their line of work.
    
    A pilot interview was done to ensure that the questions were understandable and would yield the required information~\cite{vanTeijlingen2002TheStudies.}. The pilot verified that the questions were understandable, the only change that was made was the rephrasing of a few questions. After the pilot, the interview questions were sent to the remaining participants. This could help them find a use case to talk about during the interview. Since the interview protocol did not change significantly, the data from the pilot was added to the rest of the data from the interviews.
    
    \subsubsection{Interview protocol} \label{Interview protocol}
    
    The interview started with a short introduction, where the overall structure of the interview and the goal of the study was explained. This was followed by open questions regarding the organisation the participant works for and the function of the participant withing the organisation. Next, a short explanation was given about eFLINT and the prototype that is being implemented. After this, in depth open questions were asked about the themes that were mentioned in section \ref{pattern} and about the needs and requirements for the interface. This included questions about the people involved with a use case, the different types of use cases that occur within the company, what information is needed for a use case, how this information is acquired, the different actions that are taken during a use case, the level of transparency towards the client and how a company deals with conflict resolution. During the interview, participants were encouraged to give as much in depth answers as possible by asking follow up questions. The protocol for the interviews can be found in appendix \ref{A: Interview Protocol}. The interviews were conducted in Dutch, for readability the protocol was translated into English. 
    
    \subsection{Analyzing the interview data}
    The interviews were analyzed using a thematic analysis approach~\cite{Braun2006UsingPsychology}. First the interviews were transcribed\footnote{Transcript available upon request.} and after reading and re-reading the data, initial codes were linked to interesting features. The codes were sorted into the established themes from section \ref{pattern}, if needed themes were merged or split~\cite{Cope2016OrganizingData}. Separate themes were created to reason about the interface and its functionality since that was not covered in the themes from section \ref{pattern}. This was a recursive process with back and forth movement throughout the different steps.
    
\section{Results: Establishing requirements} \label{Results: Establishing requirements}
    This section will present the results from the first phase of the research. This includes the needs, requirements and use cases that were acquired from the interviews. The results of the thematic analysis are presented in thematic maps and explained in more detail in this section.

    \subsection{Involved people} \label{Results: involved people}
    
    The involved people could be divided into three categories: clients, internal parties and external parties (figure \ref{fig: TM involved people}). Clients can be civilians, organisations or governments. Internal parties include all involved people from the organisation that take on the case. Here, there are often multiple people involved. One reason for this is to apply the four eyes principle, meaning that two individuals approve some action before it can be taken. Another reason is that in some organisations the different steps of a case are handled by different departments. Lastly, a supervisor is often involved. A supervisor usually overseas multiple cases at the same time.
    
    External parties refer to people outside the organisation. This can include people from court, for example if there is a conflict about an interpretation of a norm. It can also include people from different organisations. If a organisation needs to perform a check on someone for criminal activity, they might ask the police for information.
    
    The different people involved with a case do not all have the same level of authority, meaning that they do not all have access to the same information and some actions might be only performed by people that have the right authority.
    
    \subsection{Types of cases} \label{Results: case type}
\begin{figure}[h]
    \begin{subfigure}[b]{0.4\textwidth}
    \centering
        \includegraphics[width=\textwidth]{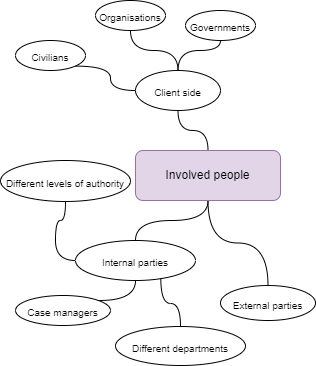}
        \caption{Thematic map: Involved people}
        \label{fig: TM involved people}
    \end{subfigure}
    \hfill
    \begin{subfigure}[b]{0.5\textwidth}
    \centering
        \includegraphics[width=\textwidth]{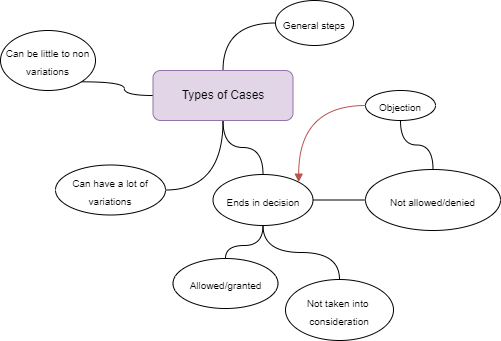}
        \caption{Thematic map: Types of cases}
        \label{fig: TM types of cases}
    \end{subfigure}
    
    \begin{subfigure}[b]{0.4\textwidth}
    \centering
        \includegraphics[width=\textwidth]{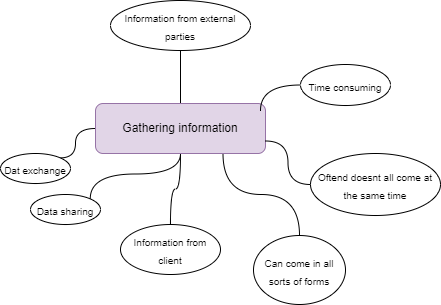}
        \caption{Thematic map: Gathering information.}
        \label{fig: TM Gathering information}
    \end{subfigure}
    \hfill
    \begin{subfigure}[b]{0.5\textwidth}
    \centering
        \includegraphics[width=\textwidth]{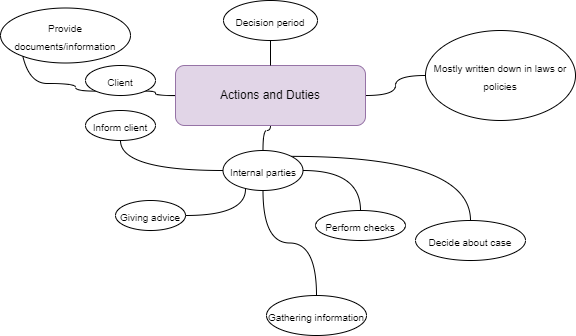}
        \caption{Thematic map: Actions and duties}
        \label{fig: TM Actions and duties}
    \end{subfigure}
\end{figure}

    This theme determined the different types of cases that can occur, specifically, the different types of workflow that occur (figure \ref{fig: TM types of cases}). The most general workflow can be broken down in a few steps: 1) request or application comes in 2) decision is made based on gathered information 3) decision is reported back to the client. How use cases fit in those three steps can depend on multiple factors. One factor was the type of client, whether the client was a civilian or a organisation can influence what information needs to be gathered in order to make a decision about the request. Another factor was what type of request is put in. For example, an objection had a different process than a request. How many different types of cases occurred varied among the different participants. One participant indicated that he only gets requests for allowance and the only thing that changed was the height of the allowance. Others however, indicated that they can have hundreds of different processes, each with slightly different steps, which made it difficult to categorize them.
    
    This theme also includes the different ways a case can end. All participants indicated that a case always ends with a decision. A request can be either, approved, denied or not taken into account. In case of the latter, the case is not accepted and does not start to begin with. If a request is approved, then the case ends there. If a request is denied, the case ends there, unless objection is made. If that happens, steps will be taken to solve the objection, but the case will then again end in a decision, this is also shown in figure \ref{fig: TM types of cases} with the red arrow that indicates the loop.
    
    Participants also reported to be interested in a simulation environment. Two types of simulation were mentioned: The first type was to see what would be the effect on a scenario if a law would change or would be added. The second type of simulation would be to see what the effect on a scenario would be if a certain action is taken. This is partly because government organisations often have to introduce a new law or a changed law. It seemed interesting to them to know how a new or changed law influences other applicable norms or scenarios.
    
    \subsection{Gathering information} \label{Results: Gathering information}
    This theme answers questions related to gathering information during a case (figure \ref{fig: TM Gathering information}). For this study the types of information were divided into two categories: information that comes from norms, laws or policies, and information that comes from clients or other external sources. Information comes in all sorts of formats, some participants indicated that they only accept information through an online form, others indicated that they can get information through online forms, e-mail, phone calls, paper documents or even one-on-one conversations. Most participants indicated that no matter in what shape or form the information is received, it should be stored. One participant mentioned that 
    
    \textit{"Information from a case should be stored in such a way that someone who is not working on that case should still be able to understand the actions that have been taken and why."} 
    
    Only one participant mentioned that he would rather use the zero-knowledge protocol~\cite{Feige1988Zero-knowledgeIdentity}. Information from finished cases is often stored, but most participants indicated that they were not sure for how long or who had access to it.
    
    \subsection{Actions and duties} \label{Results: Actions}
    This theme covers the actions that occur during the various use cases (figure \ref{fig: TM Actions and duties}). The actions and duties are divided into two categories, they are either for the internal parties or for the client/external parties.
    
    The client has the duty to provide requested information. If this duty is not fulfilled, a decision could be made with incomplete information, which usually not in favour of the client. The decision officer on the other hand, has the duty to make a decision, and sometimes do this within a given decision period or even target term. Other actions and duties varied per case type, but were often written down in related laws and policies. These actions can vary from informing a client about the decision to sending letters to other organisations with a request for information.
    
    \subsection{Interface requirements}
\begin{figure}[htb!]
    \centering
    \includegraphics[width=\textwidth]{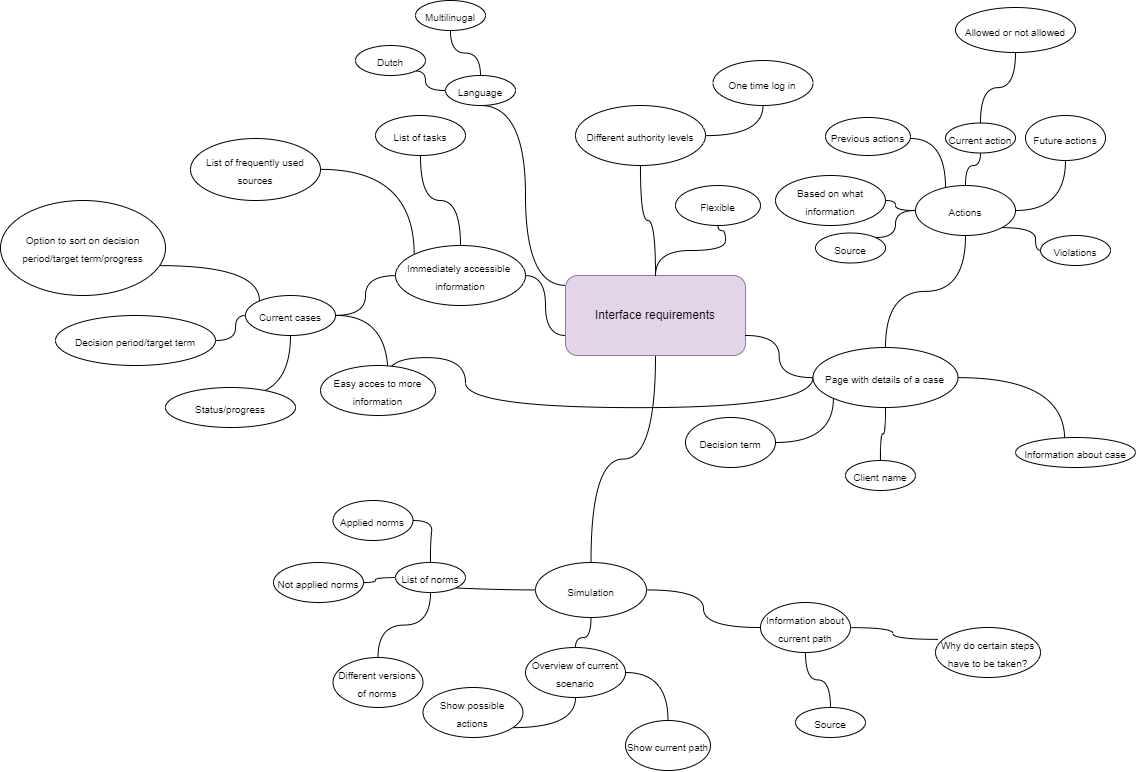}
    \caption{Thematic map: Interface requirements}
    \label{fig: TM interface requirements}
\end{figure}

    Lastly the thematic map of the requirements for the interface is presented in figure \ref{fig: TM interface requirements}. The different interface requirements were divided into different sub themes. First there were general requirements, such as the option to have different authority levels, language preference, or color blind friendly. The other themes are: \textit{Immediately accessible information}, \textit{Page with details of a case} and \textit{Simulation}. These categories will be discussed in this section
    
    \subsubsection{Immediately accessible information}
    Most participants indicated that there were certain pieces of information they want to have immediate access to. Most importantly is the cases they are currently working on, or depending on their function, certain norms they are working on. Since cases can be bound to decision terms or target terms, it is important that the user is made aware of approaching deadlines. Most participants indicated that in the information system they are currently using, they have a traffic light system to indicate which cases near their decision term, this is something they also wanted to see in the prototype.
    
    Since participants mentioned they are sometimes working on a lot of cases at the same time, sorting and filtering features are desirable. Participants also indicated they wanted to be able to see what the status of a case is and what their tasks are. As one participant mentioned 
    
    \say{\textit{I want to see all the cases I am working on while having the option to zoom in on specific parts.}}
    
    Almost all participants emphasized the need to have a link to relevant sources when working on a case.
    
    \subsubsection{Page with details of a case}
    The information about a specific case was divided into two sub-themes. The first one is information regarding the steps taken during the development of a case. The second theme is all other information. Information about the steps taken during the development of a case contains information such as: Actions that need to be taken, actions that have been taken, on what information or sources those actions are based, other relevant sources, whether actions are allowed or not and if any violation has occurred. One participant indicated that 
    
    \textit{"There is a lack of knowledge about the relevant norms and how they need to be applied, this leads to mistakes or cases not being completed on time."}
    
    The participant emphasized the importance of being clear about which laws apply and how that translates to actions that need to be taken.
    
    The other sub-theme, which contains all other information, includes information such as: client name, decision term and any other information that was gathered for a case. This also includes information delivered by the client or external parties. Since most participants indicated that information often does not all come at once, there should be an option to add information later on.

\section{Experimental setup: Implementing and testing the prototype} \label{ES: implementation and testing prototype}

This section explains the method mentioned in section \ref{Methodology}, specifically the part about implementing and testing the prototype. The steps discussed in this section are: interface design, prototype implementation, and evaluating the prototype.
    
    \subsection{Interface design}
    This section will elaborate how the established requirements were incorporated into the interface. The design can be seen in figures \ref{fig:Design home} - \ref{fig:Design simulation} in the appendix. The following sections describe each page in more detail. The target group of the prototype is Dutch government organisations, therefore one of the requirements was that the interface should be in Dutch. Some participants also indicated they they would like a multilingual interface, but Dutch was the most preferred. In this report the Dutch words and terms in the design and prototype will be used. The design of the interface was made using Framer\footnote{\url{https://www.framer.com/}}.

    \subsubsection{Home page} \label{Home page}
        The first and most important page is the \textit{Home} page (figure \ref{fig:Design home}). The \textit{Home} page holds the information that needs to be immediately accessible. Participants indicated that it is important for them to have immediate access to the cases they are currently working on. Therefore, the \textit{Home} page includes an overview of the ongoing cases, this can be found under the header \textit{Overzicht lopende zaken} (overview ongoing cases). Each row in the overview represents one case. There are three columns that hold information about the case, the last column holds the button \textit{Open} that leads to a new page, with more detailed information about the case. The first column, \textit{Naam} (Name), displays the name of the client. \textit{Status} presents the status of the case. The \textit{Termijn} (Term) presents the date for which a decision needs to be made. An added feature here is the little clock next to the date, a yellow clock means the deadline is getting near, a red, filled clock means the deadline is very near. One of the participants mentioned to be colorblind. Therefore, throughout the design an effort was made to not only rely on colors, but add different textures as well. An example is the clocks used here. The overview has a sorting function in the top right corner where the cases can be sorted on \textit{Termijn} or \textit{Actie} (Action), descending or ascending.
        
        The next piece of information on the \textit{Home} page is the list of \textit{Openstaande Acties} (Available actions), which presents the actions the user need to perform, this can also be seen as a to-do list. The difference between \textit{Lopende zaken} amd \textit{Openstaande acties} is with cases where the user has to wait for an action from another party. For example, for case \#2 the status is \textit{Wachten op bericht} (waiting for information), meaning the user is waiting for information from the client or an external party before further action, therefor this case is not shown in the \textit{Openstaande Acties} list. This way the user has an overview of the tasks he needs to perform, while also being able to keep an eye on cases where another person might have to take an action first. Each block in \textit{Openstaande acties} represents an action for a case. For each action, the name of the client, the action and the term displayed. Each block also has an \textit{Open} button to go to the more detailed pages of the case.
        
        The last piece of information on the \textit{Home} page is the \textit{Bronnen} (sources) list. Here the user can find frequently used sources. These can either be added manually or perhaps even automatically selected by the model. For each source, the name, date of applicability and link to the source are included.
        
        At the top of the page is the navigation menu, the design has three tabs, \textit{Home}, \textit{Overzicht klanten} (Overview clients) and \textit{Simulatie} (Simulation). There is also a button in the top left corner, \textit{Nieuwe zaak} (new case), here the user can create a new case.

    \subsubsection{New case page} \label{New case page}
        When the user clicks the \textit{Nieuwe zaak} button it will lead to the \textit{Nieuwe zaak} page. Here the user can fill in the information needed to start a case. On the left is the administrative information, here the user can fill in the \textit{Naam klant} (Name of client) field, the \textit{Aanmaakdatum} (Creation date), \textit{Beslistermijn} (Decision term), choose a type of request and there is a field for notes. On the right is a list of questions that need to be filled in for the model to decide what actions need to be done. The questions here are just placeholders, they were later replaced with ones more fitting to a specific use case. When everything is filled in the user can click the \textit{Volgende} (next) button and it will lead to the overview page of the case. If the user presses the \textit{Volgende} button without filling in all the administrative information, it will raise an error and ask the user to fill in the empty fields.

    \subsubsection{Case information page} \label{ Case information page}
        In the case overview page the user can find all information about the selected case. On the left, the information that was filled in when creating the case is presented. In addition, the \textit{Laatst gewijzig} (last changed) date is also shown. The user can press the pencil icon to make changes to the information, this will be explained in more detail in section \ref{Edit case page}. On the right, a list of actions is presented. The actions are divided into two lists, the top one with the header \textit{Afgerond} (Completed), holds all actions from the case that are completed. The bottom list with the header \textit{Vervolg} (follow-up) holds all possible actions that still might need to be executed. For each action, it will show the name of the actions, the \textit{Normatieve status} (Legal status), \textit{Relevante bronnen} (relevant sources), if needed, and a \textit{Uitvoeren} (execute) button to complete the action. The \textit{Normatieve status} can either be \textit{NIET toegestaan} (Not allowed), \textit{Toegestaan} (Allowed) or \textit{Onbestemd} (indefinite). If the legal status is not allowed, the \textit{Uitvoeren} button will have a red border, it will also presents information on why the action is not allowed. Both allowed and not allowed actions are executable, but for a not allowed action a motivation must be provided and the prototype should give a warning. When an action is completed it will move to the \textit{Afgerond} list. If an action that was not allowed is completed, it will get a red warning sign next to its name to indicate a violation has occurred. The violated action can be expanded to show more information about the violation.
        
    \subsubsection{Edit case page} \label{Edit case page}
        When the user clicks on the pencil icon on the case information page, he will be led to the edit case page where the information about the case can be changed. The screen looks similar to the new case page, only here the information is already filled in. After the user has made the desired changes, he can press the save button and will go back to the case overview page. When the information is changed, the legal status of actions can change, this will be visible in the case overview page. Since the edit case page is similar to the create case page, there was no separate page included in the design.
    
    \subsubsection{Cases overview page} \label{cases overview page}
        Clicking on the \textit{Zaken} button in the navigation bar leads to the \textit{Overzicht klanten} page. It is similar to the \textit{Lopende zaken} overview in the dashboard, with the difference that the case overview page contains more columns and sort, filter and search functions. The page contains two extra columns: \textit{Type} and \textit{Gewijzigd op}. The former presenting the category type of a case, the latter presenting the date the case was last changed. Although not all shown in the design, starting in the top left corner, the idea was that the user would be able to  sort on \textit{Naam}, \textit{Termijn}, \textit{Actie}, filter on date and have the possibility to use a search bar to find a specific name. Here, the \textit{Termijn} column also contains the clocks mentioned in section \ref{Home page}.
        
    \subsubsection{Simulation page} \label{Simulation page}
        The last page is the tab \textit{Simulatie}. On this page the user can create scenarios, or use existing ones, and experiment with changing the applied norms to see how that would affect the scenario. Since this is a feature that is fairly complex to implement in eFLINT, the chance that this could be included in the prototype was small. However the design was made in case a hard-coded version would be included in the prototype. Therefore the design is not as detailed as when this feature would have actually be implemented.
    
        The page is divided in three parts. On the left, a list of rules that are being used is presented, each line is one rule. If the rule is green and has a check-mark, it means it is active, if the rule is grey and has a minus sign, it means it is inactive. Each rule has a pencil icon, which can be used to edit the rule, and a plus sign. The plus sign is used to add a different version of the rule. If there are multiple versions of one rule, they will appear under the rule that they originate from. The active version will be highlighted in blue and will have the check-mark. Rules, or versions of rules, can be deleted by filling the checkbox on the left and clicking the trashcan located above the list of rules. There are four more buttons above the list of rules. Starting in the top left corner, the button \textit{Regel toevoegen} (Add rule) can be used to add a new rule. The grey button next to it, \textit{Deactiveer alles} (deactivate all) can be use to make all rules inactive. The blue button in the bottom left corner, \textit{Selecteer alles} (select all), can be used to fill the checkboxes from all rules. Lastly, the green button in the bottom right, can be used to activate all rules.
        
        The next part is the middle part. In the middle part a tree-like structure presents what the process of completing a case would look like with the applied rules. Each circle represents an action, if the circle is green it means it is selected, if the circle is grey it means the action is not selected. The user can click on the circles to choose between the different paths that are available, which will then be highlighted in green. Each circle contains a label with the name of the action. If rules are activated or deactivated, the three structure will change accordingly.
        
        On the right side of the screen is a list with all the highlighted actions from the middle part. Here each action can be expanded to provide more information e.g., about what rules apply to this specific action. This list can also be fold in to make more space for the tree structure in the middle by pressing the arrow on the top left of the list.

    \subsection{Prototype implementation} \label{The prototype}
    
     The prototype\footnote{\url{https://eflint-prototype.nl/register}} itself was implemented by someone else. Due to limited time and resources there were some differences between the design and the prototype that was actually implemented, some more significant than others. The next section  discusses the prototype and some of the differences with regard to the design. 
     
     \subsubsection{IIT use case} \label{IIT use case}
     To give some context to the prototype, one of the use cases from the interviews regarding the \textit{Individuele Inkomenstoeslag (IIT)} (Individual income allowance) from the municipality of Utrecht was worked out in more detail in eFLINT. A few changes were made to the use case to better represent the role of eFLINT. The use case provided was about IIT, the idea is that a civilian can request IIT by sending in their information, the decision officer fills in the information in the prototype and based on that the model presents what actions should be taken accordingly. This means that the model decides whether the civilian gets the IIT and how much. This decision however, can be overruled by the user. The model is based on the rules specified in an example from \url{open-regels.nl}~\cite{Open-regels.nl}. Since the focus is on how eFLINT can help the decision officer while processing such request, how exactly the information gets in the system was not taken in account and was done manually. 
    
    \subsubsection{Differences between prototype and design}
    The first difference between the design and the prototype was the log in screen. The log in screen is the first screen the user sees when launching the prototype. Here the user can either log in or register. This feature was added since interviews indicated that there are often multiple people involved with a case with different levels of authorization. However, the prototype currently only has one level of authorization, meaning that anyone who logs in can see all information.
    
    Another significant difference was the simulation tab. Due to time constraints and the complexity the system would have, it was decided to not include the simulation tab in the prototype.
    
    Lastly, in the case overview page the completed actions did not show any information. As this was not feasible to implement within the given time limit. Not allowed actions did not give a warning when executed, did not require a motivation and did not show why they are not allowed.

    \subsection{Prototype evaluation} \label{Evaluation}
    
    In this section the set up for the evaluation is discussed. The evaluation was done through expert evaluation and user testing. Expert evaluation involved a cognitive walkthrough. User testing consisted of usability testing along with interview questions. From the seven participants, six did the user test. 
    
    \subsubsection{Cognitive Walkthrough} \label{cognitive walkthrough}
    The first step for the cognitive walkthrough was to set goals to be accomplished with the prototype~\cite{Rieman1995UsabilityWalkthrough}. These goals were based on the use case from section \ref{IIT use case}. Next, the sub-tasks needed to complete the goals were established. The cognitive walkthrough was performed with the researcher taking on the role as the expert. The expert performed the set goals with the selected actions and evaluated the system. This was done based on three questions for each sub-task: "Will the user know what to do to achieve the task?", "Can the user see the button/menu item needed to achieve the task?" and "Will the user know from the feedback whether the action was correct (and what to do if not)?". These questions were answered with a simple yes or no for each sub-task and notes were taken if needed. The complete cognitive walkthrough can be found in appendix \ref{A: Cognitive walkthrough}.
    
    According to the cognitive walkthrough the users should not have to much trouble completing the tasks. There were three remarks made. The first one being the goal where the user has to check if a violation has occurred in a specific case. There was a warning sign that indicates a violation. However, the warning sign did not give feedback which might confuse the users. The second remark was that users have multiple options to open a case, either through the overview in the dashboard, the list of actions in the dashboard or through the \textit{Zaken} tab. It would be interesting to see during the user test if there is a specific way they prefer. The last remark was when the user was asked to change a piece of information from a specific case which alters which actions are allowed. Users might not notice straight away that changing the information also changed the actions. Attention should be payed to this by the researcher during the user testing.
    
    \subsubsection{User testing}
    The user testing started with a short introduction. Here it was made clear again that the interface was for a generic prototype, rather than for a single use case. This to prevent mismatch in expectations between use case providers and the designed interface. 

    After the introduction, the participants were asked to complete the goals established during the cognitive walkthrough. The participants were only told what the goal was, not the sub-steps or how they could complete the goal. During the execution of the tasks, the participants got minimal to no guidance, as the interface should be intuitive and easy to use and therefore need no extra explanation. After the usability test, the user was asked to give feedback on the prototype through a series of questions.
    Here, the participants had time to point out any features that they thought were usefull, could be left out, or might be missing. The user test protocol can be found in appendix \ref{A: User Test Protocol}
    
    The user tests were also recorded. The participants were asked to use the thinking aloud method, this helped the process of understanding how they experienced the prototype and why they completed tasks the way they did.

\section{Results: Testing the prototype} \label{Results: Testing the prototype}

This section presents the results of the evaluation of the prototype. First, the results of the user tests are presented. Next, the results from the interviews following the user tests are discussed.

    \subsection{User test Results} \label{User test results}
    
    In this section the results from the user tests are presented. As explained in section \ref{Experimental Setup}, the user tests consisted of two parts. First was the usability test, followed by interview questions.
    
        \subsubsection{Usability test} \label{usability test}
        After the first user test it turned out that the prototype was not developed enough to get useful information from the user test. Therefore it was decided to push back the other user tests and spend a bit more time on developing the prototype. Because of this, the first user test was done with a different version of the prototype. The participant from the first user test did another user test with the new version of the prototype. The results form the user tests will be discussed by going over the goals one by one. 
        
        \textbf{Goal 1:} \textit{Create a new case.} Here the users were given information they could use to create a new case. Some users had difficulty with finding the client name field. The prototype does give a warning if the user tries to proceed without filling in the required fields, but a few users needed to be shown where the client name field was. The provided information did not contain a decision term, this resulted in some users not filling in the \textit{Beslistermijn} field. Which, again, led to the prototype giving a warning message. 
        
        The information provided was meant to ensure the user ended with a case with an allowed action. However, one of the users pointed out that although the information said the civilian lived in Utrecht, that did not mean the civilian was registered in the municipality, therefore the user filled in \textit{Niet bekend} (unknown) for the corresponding question. This lead to the model not allowing any actions, which limited completion of other goals. Therefore, after completing the goal, the user was asked to change the information from the case accordingly. After this, the usability test proceeded as usual.
        
        \textbf{Goal 2:} \textit{Check whether there has been a violation in case UserTest1.} This goal turned out to be difficult to complete. As mentioned in section \ref{cognitive walkthrough}, since the warning sign did not provide feedback, the users were not sure whether they completed the goal or not. As one of the participant mentioned: 
        
        \say{\textit{I thought it had something to do with the warning sing, but when I clicked on it, nothing happened. I figured I needed to look for something else.}}
        
        \textbf{Goal 3:} \textit{Change the information from the case just created, now the civilian does not have a child living at home.} For this goal the users had to change information from the case. All users were able to complete this tasks, however, not all of them noticed that changing the information also changed which actions were allowed.
        
        \textbf{Goal 4:} \textit{Execute the action from the case with the nearest deadline.} This goal was also completed by all users. What stood out is that most users went to the \textit{Zaken} page and instinctively pressed on the \textit{Termijn} column to try and sort it on decision date. This function was however not added yet to the prototype. Interesting enough, none of the users used the \textit{Openstaande acties} list, where the first action stood at the top of the list.

        \subsubsection{Interview about the  prototype}
        One of the first things most participant mentioned is that they liked the clean design of the prototype. They mentioned that except for last part of goal 2, it was clear what they needed to do and how they could complete the goal. 
        
        Starting with the dashboard page, from the three features (the \textit{Lopende zaken} overview, the \textit{Openstaande acties} list and the \textit{Bronnen} list), the \textit{Lopende zaken} was the only used feature. Users did hover over the \textit{Openstaande acties} list whilst scanning the page, but when they needed to open a case, they all either used the \textit{Lopende zaken} overview or the \textit{Zaken} tab. The participants were asked during the interview if they understood the purpose of the \textit{Openstaande acties} list. The answers varied. Most participants seemed to have the right intuiting, meaning that they thought the list was some sort of to do list. However, none of the participants could figure out what the exact difference was between \textit{Openstaande acties} and \textit{Lopende zaken}. After the researcher explained the idea behind the list, most participants agreed that it could be a useful addition, but it would need to be made more distinct. One of the interviewers mentioned:
        \say{\textit{Instead of the overview I would like to only have the list of actions and then keep the overview on a separate page.}}
        
        Almost all participants liked the feature of the colored clocks in the \textit{Lopende zaken} overview, only most participants did think the red clock meant the deadline already passed, instead of the deadline being very close. Only one participant indicated that the clocks did not stand out enough, the participants suggestion was to color the whole row to make it more clear. There also seemed to be some confusion about the headers of the columns from the overview. Whether \textit{Naam} meant the name of the client or the name of the case, or why there is a date in the column from \textit{Beslistermijn} while the decision period is a period, not a date. The \textit{Bronnen} list was not given much attention during the user tests. Participants indicated during the interview that they thought it was good to have a list of frequently used sources, but that it might make more sense to have that list on the case overview page, since that is the place where they would be working.
        
        Next is the new case page. The most made comment about the new case page was that information such as a decision term or starting date should have been filled in automatically. Other feedback was mostly about minor details, such as when asking if a client has a partner, the system would usually also ask for the name of the partner or other information to verify the partner. One of the participants mentioned that a distinction should be made between information about the case and information about the client, especially since one client can have multiple cases. 
        
        Regarding the case overview page, most of the participants mentioned that more information could have been added to the case overview page, specifically information regarding the normative sources and the reasoning of the model. Adding this type of information would help the users better understand how the model works. Currently the \textit{Bronnen} list is on the dashboard page, but frequently used sources could also be located in the case overview page. Most of the participants also mentioned the lack of feedback regarding the violation.
        
        Most feedback from the participants about the \textit{Zaken} tab was similar to that from the \textit{Lopende zaken} overview, since the two are quite similar. This was also one of the comments most participants made:
        
        \say{\textit{If there are two different pages, I also expect them to have different functionalities.}}
        
        The \textit{Zaken} tab does have an extra column and (non functional) search features, but according to the participants it does not stand out enough to be useful as a separate tab.
        
        Since the prototype was not designed for a specific organisation or function, some users found it difficult to tell whether the prototype could be used within their type of work. All of them noted that the prototype was a step in the right direction, but only a few said they saw a purpose for the prototype in its current form in their work. One participant said that some of the employees lack experience and therefore knowledge needed to correctly work on the cases. The participant said a system such as the prototype could be a valuable addition to their own information system, by supporting employees and ensuring compliance of the applicable norms and policies. Most participants mentioned they would need a more developed or specific prototype in order to decide whether it could help them or not.

\section{Discussion} \label{Discussion}
In this section the results from the first and second phase of the research are discussed.

Since the participants all had different functions, it was at times difficult to find patterns within the answers they had given during the interviews. One example of this is that some participants were more interested in a simulation environment and less in a system that supported the users. Part of this could be due to the fact that for some participants, a big part of the process from their use cases was already automated which left little room for eFLINT. One theme that was particular challenging to define was the \textit{Actions and duties} theme. The actions and duties seemed to vary too much between the different participants to establish more than a few general actions and duties.

The results from the user test confirmed the remarks made during the cognitive walkthrough. When completing the goal where they had to check for a violation, all users seemed confused on whether they had completed the task or not. They indicated that they saw the warning sign and expected it had something to do with the violation, but since it did not provide feedback they were unable to tell whether they did the right action or not. The other remark that was made is that there are multiple options to open a case. All users either chose to open a case through the overview in the dashboard or by using the \textit{Zaken} tab. No one used the \textit{Openstaande acties} list. Part of this might be due to the way the goal was phrased. When the question was initially asked, users often asked if it could be repeated. This led to the researcher repeating the goal but first saying to find the specific case and then check for a violation in an attempt to make the question more clear. Since the first thing that was said then was to find case X, the users might have taken a different approach then when the first thing they heard was to check for a violation. However, during the interview part, users also indicated that most of them did not fully understood the purpose of the \textit{Openstaande acties} list until the researcher explained it.

The \textit{Bronnen} list was not part of the goals, which might be a reason as to why the users did not seem to notice it during the user test. After explaining its purpose, participants  ]said they thought it was a nice feature, but it might be in the wrong place. However, if the overview page would have contained more information about the relevant norms the \textit{Bronnen} list might still be a useful addition to the \textit{Dashboard} to find frequently used sources that are not specific to one case.

    \subsection{Limitations}
    
    Conducting interviews meant that part of the management of expectations was dependent on the agenda of the participants. This resulted in some delay in the research planning and implementation of the prototype. 
    
    One thing that needed to be taken into account was the expectations of the participants. Since the prototype that was being developed was a general prototype, it could lead to wrong expectations from the participants. To resolve this the communication about this was clear during each step of the project. Since a general prototype was implemented instead of one specific to one organisation, the resulting prototype was not immediately useful for all participants. Some were more interested in different aspects that were not covered during this research such as the simulation tab, which was not implemented.

\section{Conclusion} \label{Conclusion}


eFLINT could potentially be useful for government organisations, but its current interface prevents adaption by legal experts. The aim of this research was to make a first step in the development of an eFLINT based prototype for automating compliance in government organisations. Interviews were conducted for requirement elicitation and to get a better understanding of the use cases from potential users. Based on the requirements and use cases, a prototype was implemented. The prototype was evaluated by expert evaluation and user testing.

This research aims to answer the question regarding what interface is required for a system based on eFLINT, that is used for automating compliance of norms and policies within government organisations. The sub questions that belong to this research question are answered in the first phase of the research, where requirements and use cases were established.

Evaluation the prototype confirmed that, although some participants might need a more specified prototype in order to know whether eFLINT or the prototype could be of use for them, the participants agreed that the prototype definitely is a step in the right direction. The prototype has a lot of room to further develop and the participants were interested in further updates.

\subsection{Future work} \label{future work}
    The prototype has a lot of room to develop further. One option would be to use the feedback from this research to improve the current version and add features that were missing. This can be features such as the simulation or more transparency as to how the model makes its decisions and what laws or policies are behind it.
    
    One feature some participants indicated to be interested in is the option to be able to make changes to existing norms and see how that would affect certain scenarios. Laws change constantly and being able to see how those changes affect other norms or scenarios could be valuable. Such simulation feature could also be used to see how certain actions would affect a scenario. 
    
    Due to time constraints this was not incorporated in the prototype. However, this is something that could be interesting to work on in future work. A simulation feature could help users to cope with changes in laws and try different actions in scenarios to see what the outcome will be. Besides, it could also be used as some test environment if users want to make changes to norms, to see what the effect of a different norm would have.
    
    Another limitation here was the implementation. Implementing such a simulation system is a complex process. The first challenge would be to design an interface where all the information needed for a simulation can be clearly represented, since there might be many different norms, with different versions. Within a given situation there could also be multiple actions that can be taken, creating different paths. 
    
    With further development the question is whether the focus stays on developing a general prototype, or if it should be more focused on a prototype for a specific organisation. As mentioned earlier, it was difficult for some participants to say whether or not the prototype could be of use for them since it was not exactly tailored to their needs and requirements. Even with a general prototype, some improvements could be made that could help users see the potential of eFLINT. However, this research also showed that even within the same organisation people can have functions with completely different needs. Therefore it might be wise to start focusing on a more specified prototype when possible.

\bibliographystyle{ACM-Reference-Format}

\bibliography{references}
\newpage
\onecolumn

\newpage
\appendix

\begin{appendices}

\section{Interview Protocol} \label{A: Interview Protocol}

    \begin{itemize}
        \item \textbf{Introduction}
            \begin{itemize}
                \item Introduce myself
                \item Express gratitude for participating
                \item Explain overall structure of the interview.
                \item Make clear that the participant does not have to answer questions if they do not want to and can stop the user test at any time.
                \item Ask permission to record the interview.
                \begin{itemize}
                    \item Recordings will be transcribed and anonymized.
                \end{itemize}
                \item Are there details that cannot be used in the report?
                \item Do you have any questions before we start?
                \item Basic information
                \begin{itemize}
                    \item Name
                    \item Company name
                    \item What does your organization do?
                    \item What does your function entail?
                \end{itemize}
            \end{itemize}

        \item \textbf{Briefly explain eFLINT and the prototype}
        \item \textbf{Questions about the involved people}
            \begin{itemize}
                \item What people are involved with a case and what are their roles?
                \item What types of clients do you get?
                \begin{itemize}
                    \item E.g., companies, civilians.
                \end{itemize}
                \item What laws and organisational policies are involved with a case?
                \item (Based on those laws and policies,) what are the most important powers and duties of each stakeholder?
                \begin{itemize}
                    \item Do all involved people have access to the same data?
                    \item Do people have authority to overrule actions that are not allowed?
                \end{itemize}
            \end{itemize}
            
        \item \textbf{Questions about the types of use cases}
        \begin{itemize}
            \item What kind of use cases does your company encounter?
            \begin{itemize}
                \item Do different types of cases each follow their own flow or pattern?
            \end{itemize}
            \item How are the involved laws and organisational policies implemented in day-to-day operations?
            \begin{itemize}
                \item E.g., are there certain steps you have to take according to company policy?
            \end{itemize}
            \item Can you describe what the process of handling a case is like, looking at typical scenarios?
            \begin{itemize}
                \item Is there a certain pattern you follow (depending on the case)?
            \end{itemize}
            \item What are different ways a case can end?
            \begin{itemize}
                \item Is any data stored. If so, for how long?
            \end{itemize}
            \item What parts of handling the case are most time-consuming?
            \item Are there parts that are already automated with software? And how does this work?
            \item Which parts are ideally automated according to you?
            \item How do you deal with conflict resolution?
            \begin{itemize}
                \item For example, if another client gets involved in a case, does it become one case or do they stay as separate cases?
                \item What if there is a conflict about the interpretation of a law or policy?
            \end{itemize}
            \item Would you want to be able to adjust interpretations of laws or policies?
            \item When working on cases, are you bound to a decision period?
            \begin{itemize}
                \item If so, does this vary? Would you like to get reminders in the interface?
            \end{itemize}
        \end{itemize}
        
        \item \textbf{Questions about gathering information}
        \begin{itemize}
            \item How do you gather information about the clients that is needed for the case?
            \begin{itemize}
                \item Do they come to the organisation for a one on one talk?
                \item Online form?
                \item Is all information gathered at once or is more added later?
            \end{itemize}
            \item Are there parts of the case where the client has the responsibility to move things forward? Or is everything done under guidance of a case manager?
            \item Can the client access the information about the case?
            \begin{itemize}
                \item How much information can they see?
                \item Can they see the same amount of information as the case manager?
                \item Can they see all the sources from the actions that are taken?
            \end{itemize}
        \end{itemize}
        
        \item \textbf{Questions about actions and duties}
        \begin{itemize}
            \item What are the different types of actions or duties that might need to be done during a case?
            \item If there are duties, and consequences, who acts upon those consequences?
            \item If an action is not allowed, are there people who can overrule this decision?
            \begin{itemize}
                \item If so, who and how does that work?
                \item Do they need to give some form of motivation?
            \end{itemize}
        \end{itemize}
        
        \item \textbf{Questions about needs and requirements for the interface}
            \begin{itemize}
                \item What are things from the current process/software that you would like to change?
                \item What are things from the current process/software that you like/don not want to change?
                \item What are features that you think are important for the interface and why?
                \begin{itemize}
                    \item Overview of current cases and their status?
                    \item Overview of actions/duties that need to be done and for what case/by whom?
                    \item Database with old cases?
                \item Reminders for decision period?
                \item Log in screen
                \item Overview of a case, with all the decision that have been made, actions that need to be taken and the source?
                \item Link to the source?
                \end{itemize}
                \item What information needs to be immediately accessible?
                \item What functions do you use the most from the current software?
                \item Who can access information from a case and how is this regulated?
                \item What information needs to be saved? And how is it currently saved?
            \end{itemize}

        \item \textbf{Closing}
            \begin{itemize}
                \item Restate the main aim of the interview
                \item Do you believe there is anything important that we have not covered?
                \item Do you have any comments or questions? 
                \item Share contact info
                \item again express gratitude for participating
            \end{itemize}
    \end{itemize}

\newpage

\newpage

\newpage

\section{User Test protocol} \label{A: User Test Protocol}

\begin{itemize}
    \item \textbf{Introduction}
    \begin{itemize}
        \item Express gratitude for participation.
        \item Ask permission to record user test.
        \begin{itemize}
            \item Recordings will be transcribed and anonymized.
        \end{itemize}
        \item Explain the aim of this study
        \item Explain overall structure of the user test
        \item Express that it is a general prototype, not for a specific use case, so it might not include all their personal requirements.
        \item Make clear that the participant does not have to answer questions if they do not want to and can stop the user test at any time.
        \item Ask if the participant has any questions before we start.
    \end{itemize}
    
    \item \textbf{Usability test}
    \begin{itemize}
        \item Give a short explanation about the prototype.
        \item Ask user to think out loud when performing the tasks.
        \item Ask the user to not ask questions when performing the goals
        \item Goals
        \begin{itemize}
            \item Create a new case using the following information: Civilian, 30 years old, lives in Utrecht, single, child living at home, income:1000 wealth: 4000.
            \item Check whether there has been a violation in case \textit{UserTest1}
            \item Change the anwer from the case you just created, so now the civilian does not have a child that lives at home.
            \item Execute the action form the case with the nearest deadline.
        \end{itemize}
    \end{itemize}
    
    \item \textbf{Questions}
    \begin{itemize}
        \item Is the information that you would need accessible?
        \item Are there parts from the prototype that you found confusing?
        \item Are the parts form the prototype that you liked?
        \item Are there features you think are missing from the prototype?
        \item Are there features you think could be left out of the prototype?
        \item Do you think this prototype could be of use to you or your company?
    \end{itemize}
    
    \item \textbf{Closing}
    \begin{itemize}
        \item Restate the aim of the evaluation.
        \item Ask the participant if there is anything they want to discuss that has not been covered yet.
        \item Ask the participant if they have any other questions or comments.
        \item Again express gratitude for participating.
    \end{itemize}
\end{itemize}

\includepdf[scale = 1.1, offset=0 -90, pagecommand=\section{Cognitive Walkthrough}\label{A: Cognitive walkthrough}]{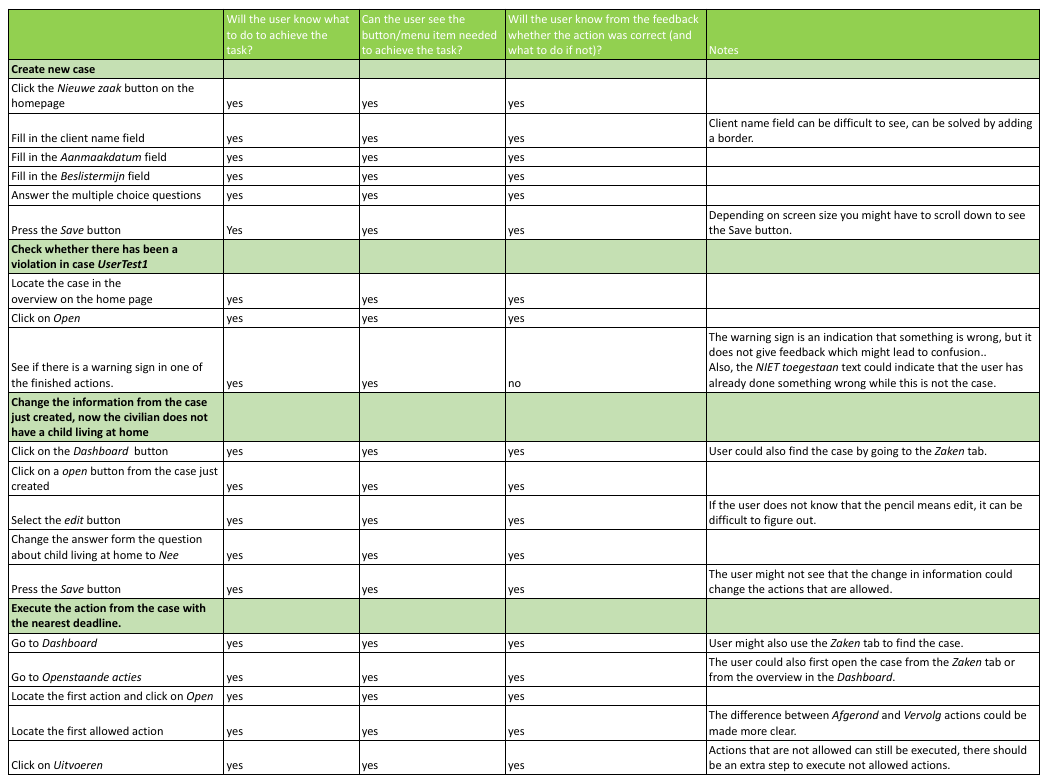}

\newpage

\section{Interface Design} \label{A: Interface Design}

\begin{figure}[htb!]
    \centering
    \setlength{\fboxsep}{0pt}
    \setlength{\fboxrule}{0.5pt}
    \fbox{\includegraphics[width=\textwidth]{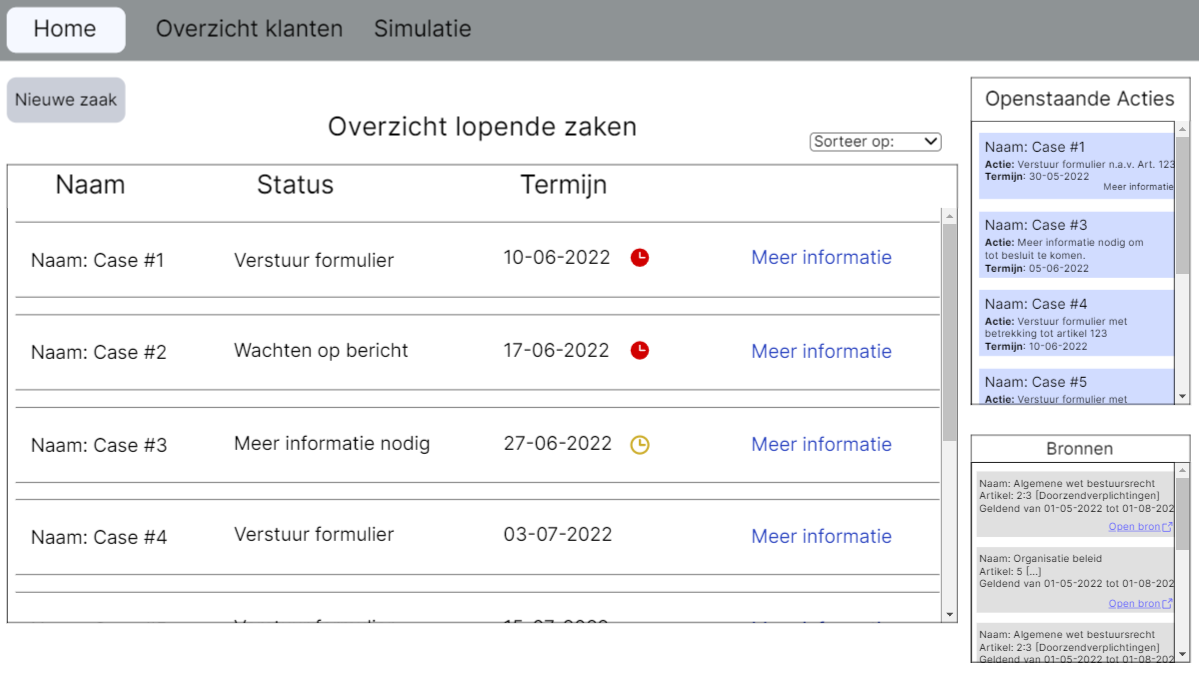}}
    \caption{Home page.}
    \label{fig:Design home}
\end{figure}

\begin{figure}[htb!]
    \centering
    \setlength{\fboxsep}{0pt}
    \setlength{\fboxrule}{0.5pt}
    \fbox{\includegraphics[width=\textwidth]{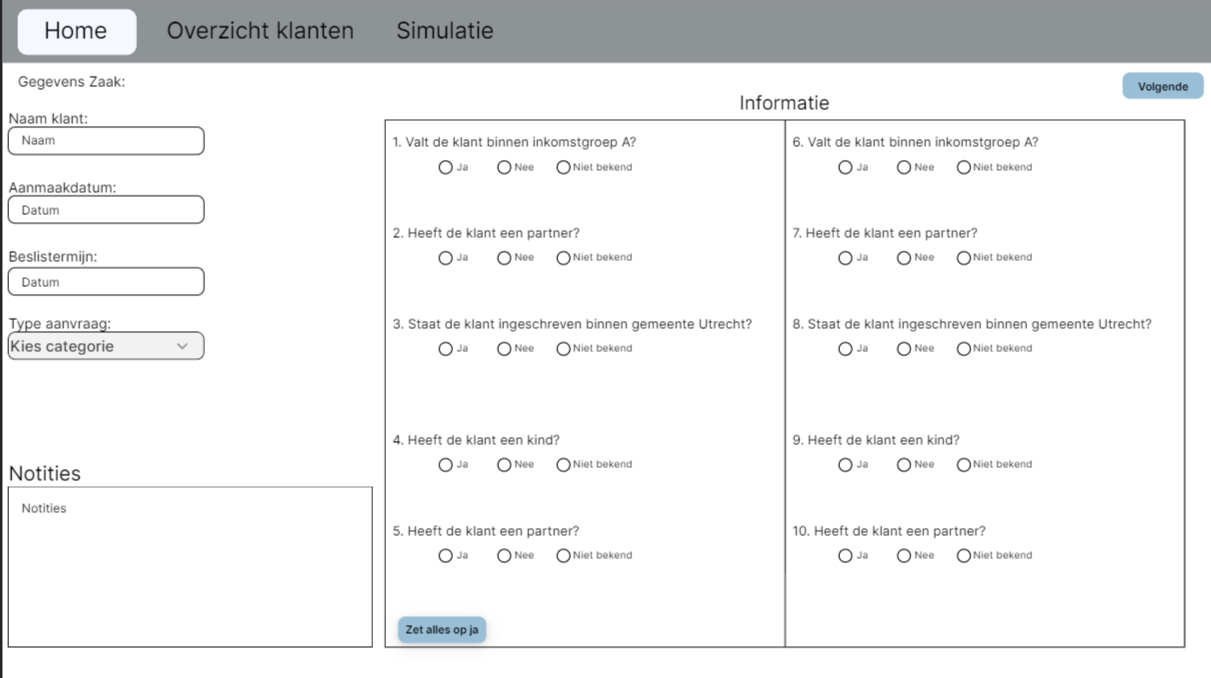}}
    \caption{New case page.}
    \label{fig:Design new case}
\end{figure}

\begin{figure}[htb!]
    \centering
    \setlength{\fboxsep}{0pt}
    \setlength{\fboxrule}{0.5pt}
    \fbox{\includegraphics[width=\textwidth]{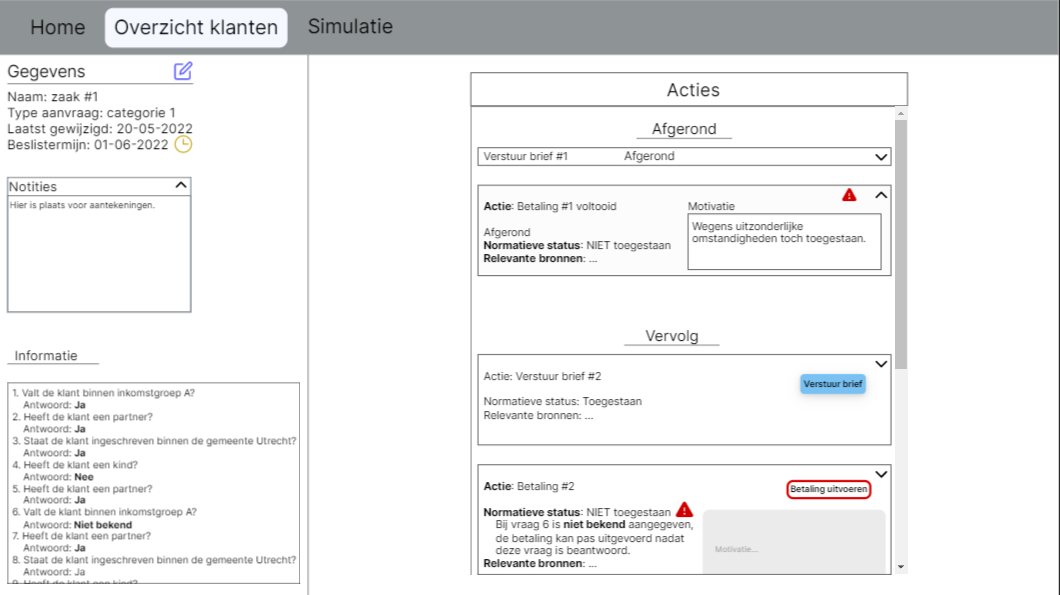}}
    \caption{Case overview page.}
    \label{fig:Design case overview}
\end{figure}

\begin{figure}[htb!]
    \centering
    \setlength{\fboxsep}{0pt}
    \setlength{\fboxrule}{0.5pt}
    \fbox{\includegraphics[width=\textwidth]{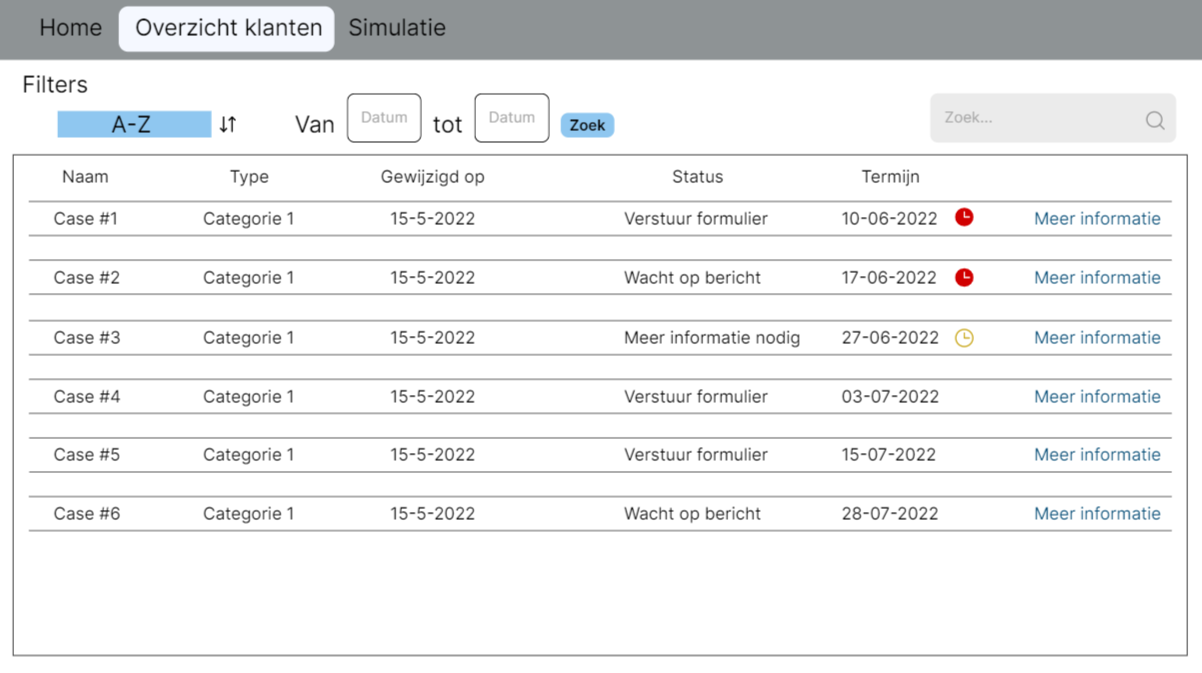}}
    \caption{Overview cases page.}
    \label{fig:Design overview cases}
\end{figure}

\begin{figure}[htb!]
    \centering
    \setlength{\fboxsep}{0pt}
    \setlength{\fboxrule}{0.5pt}
    \fbox{\includegraphics[width=\textwidth]{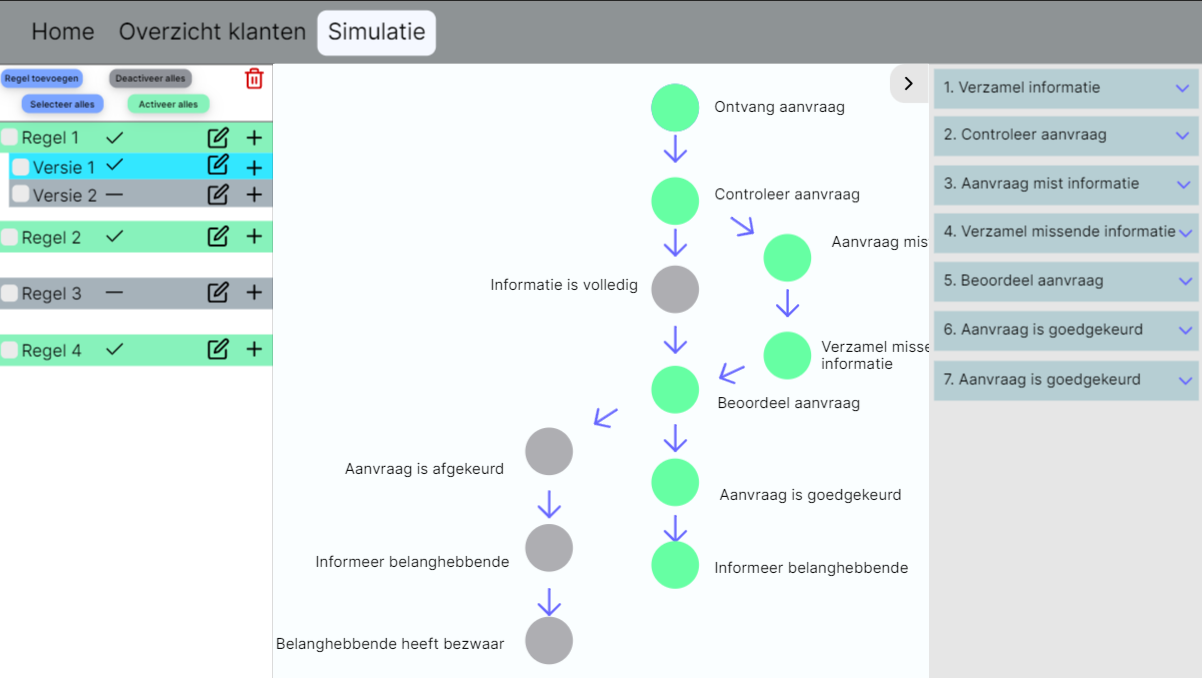}}
    \caption{Simulation page.}
    \label{fig:Design simulation}
\end{figure}

\FloatBarrier

\section{Prototype} \label{A: Prototype}

\begin{figure}[htb!]
    \centering
    \setlength{\fboxsep}{0pt}
    \setlength{\fboxrule}{0.1pt}
    \fbox{\includegraphics[width=\textwidth]{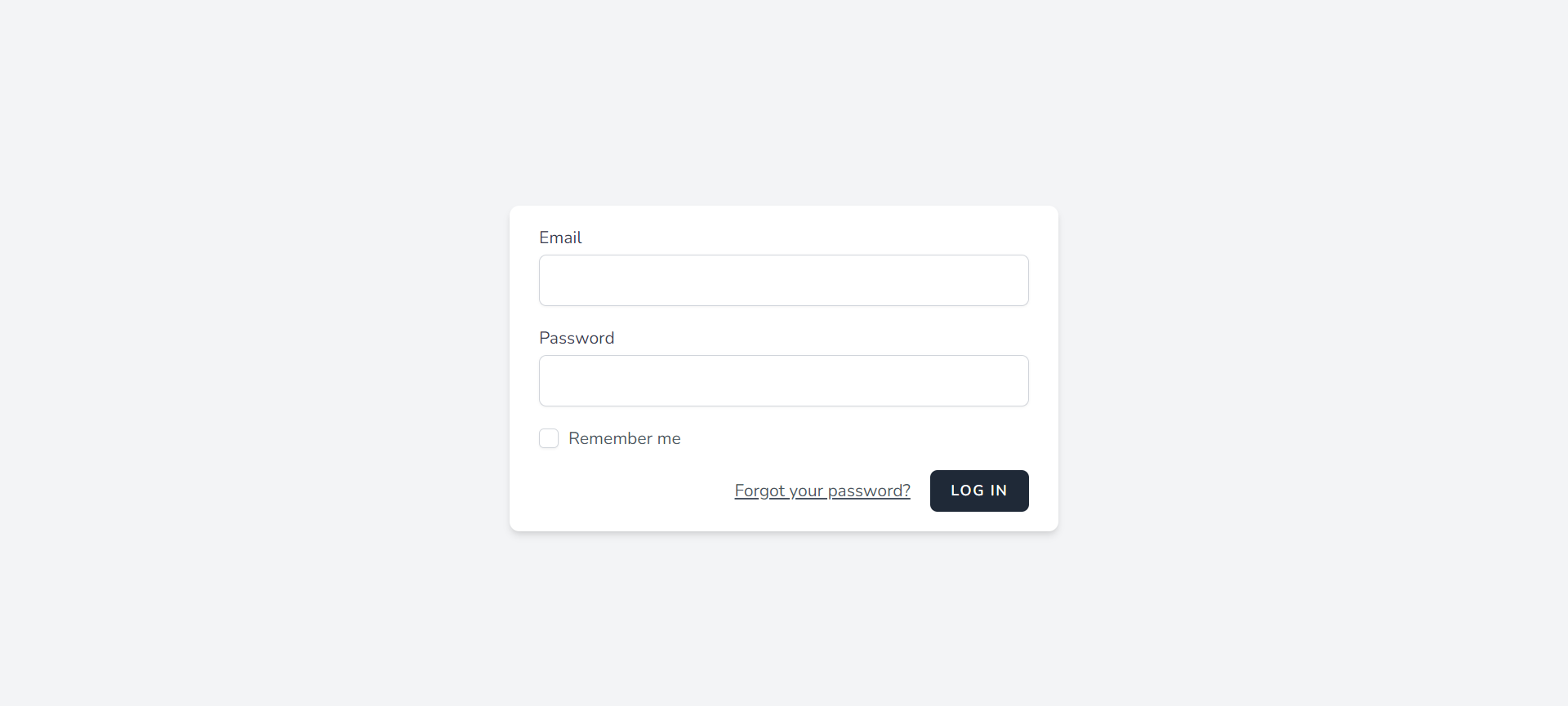}}
    \caption{Log in page of the prototype.}
    \label{fig:log in}
\end{figure}

\begin{figure}[htb!]
    \centering
    \setlength{\fboxsep}{0pt}
    \setlength{\fboxrule}{0.1pt}
    \fbox{\includegraphics[width=\textwidth]{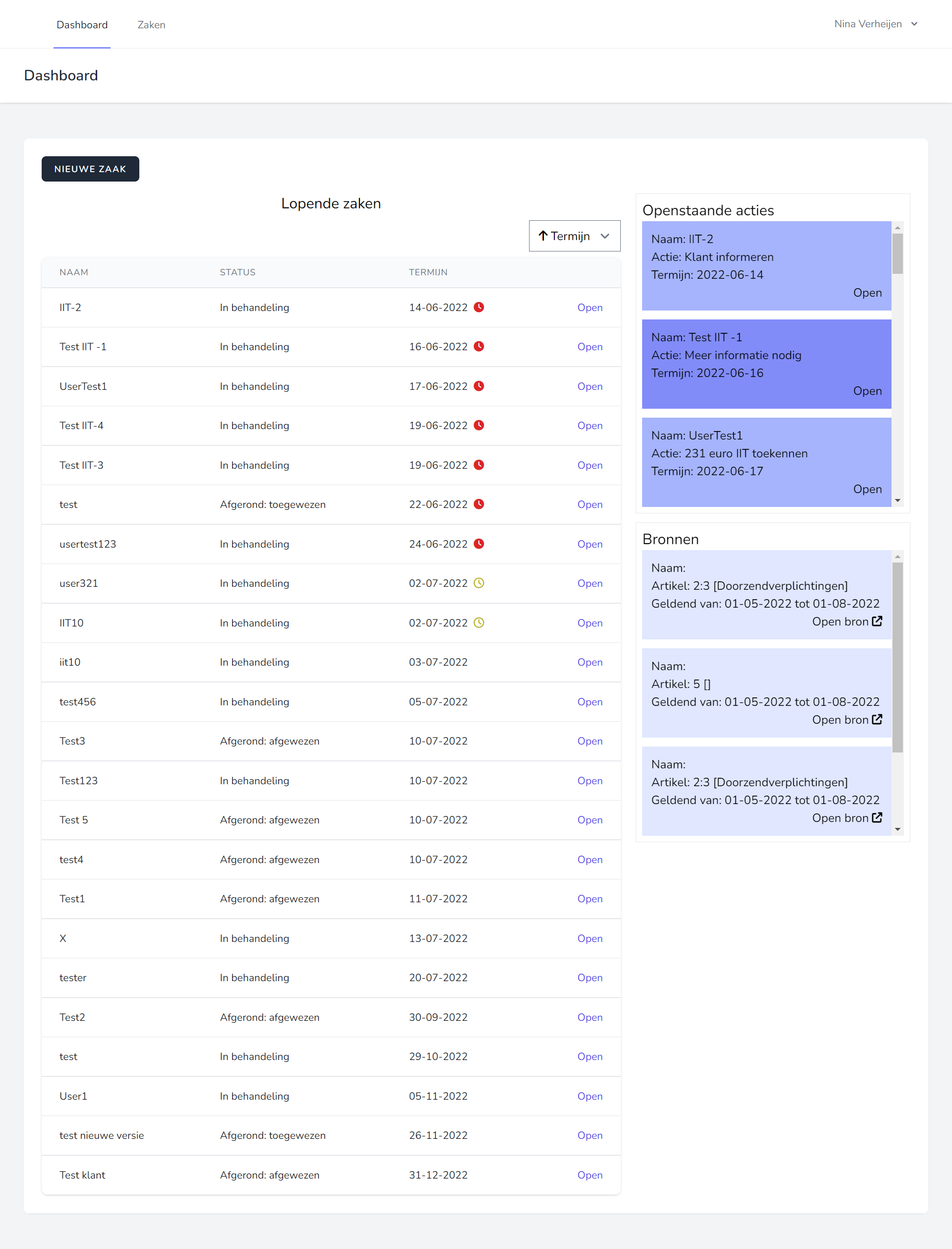}}
    \caption{Dashboard of the prototype.}
    \label{fig:dashboard}
\end{figure}

\begin{figure}[htb!]
    \centering
    \setlength{\fboxsep}{0pt}
    \setlength{\fboxrule}{0.1pt}
    \fbox{\includegraphics[width=\textwidth]{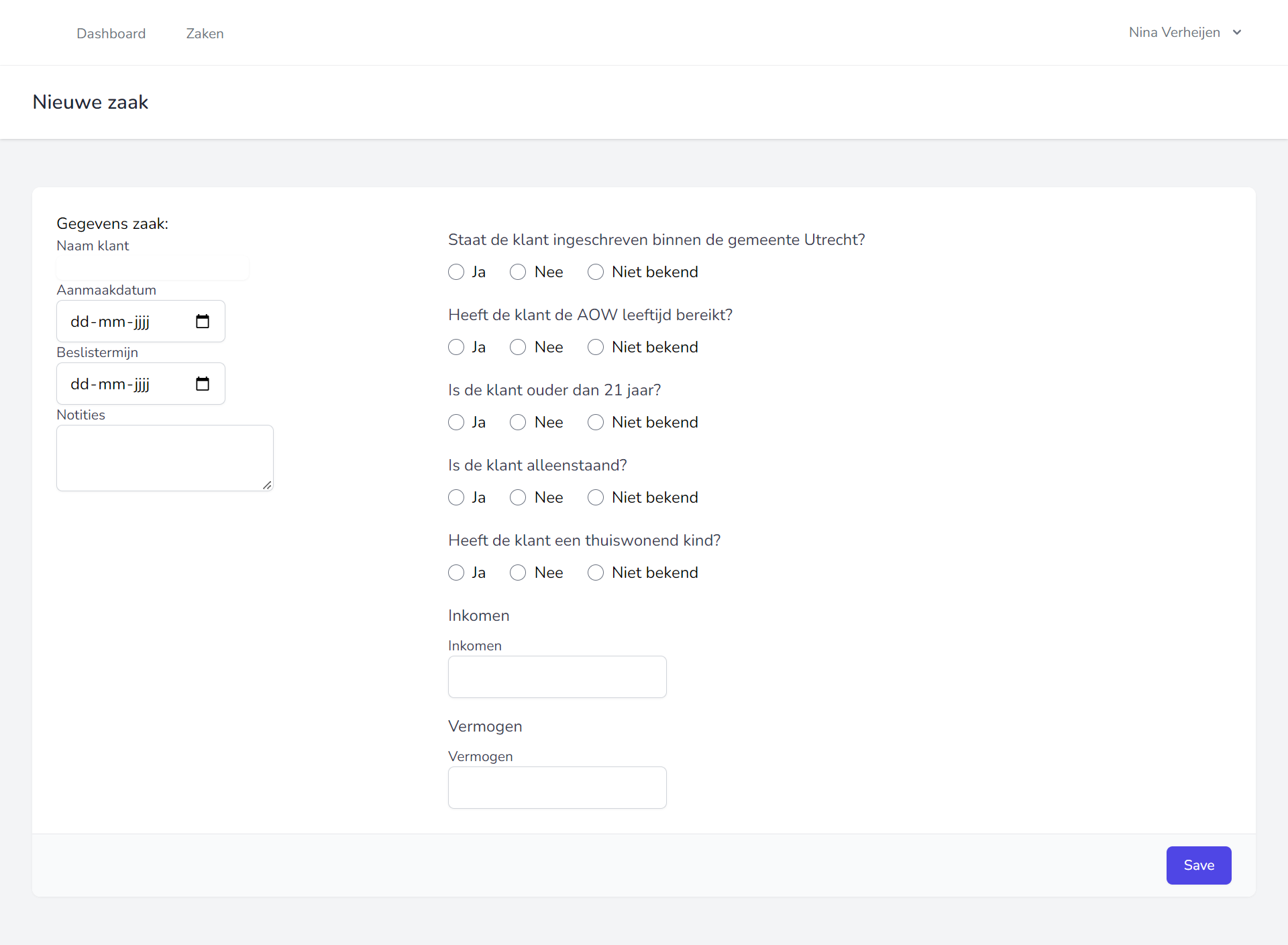}}
    \caption{Create case page of the prototype.}
    \label{fig:create case}
\end{figure}

\begin{figure}[htb!]
    \centering
    \setlength{\fboxsep}{0pt}
    \setlength{\fboxrule}{0.1pt}
    \fbox{\includegraphics[width=\textwidth]{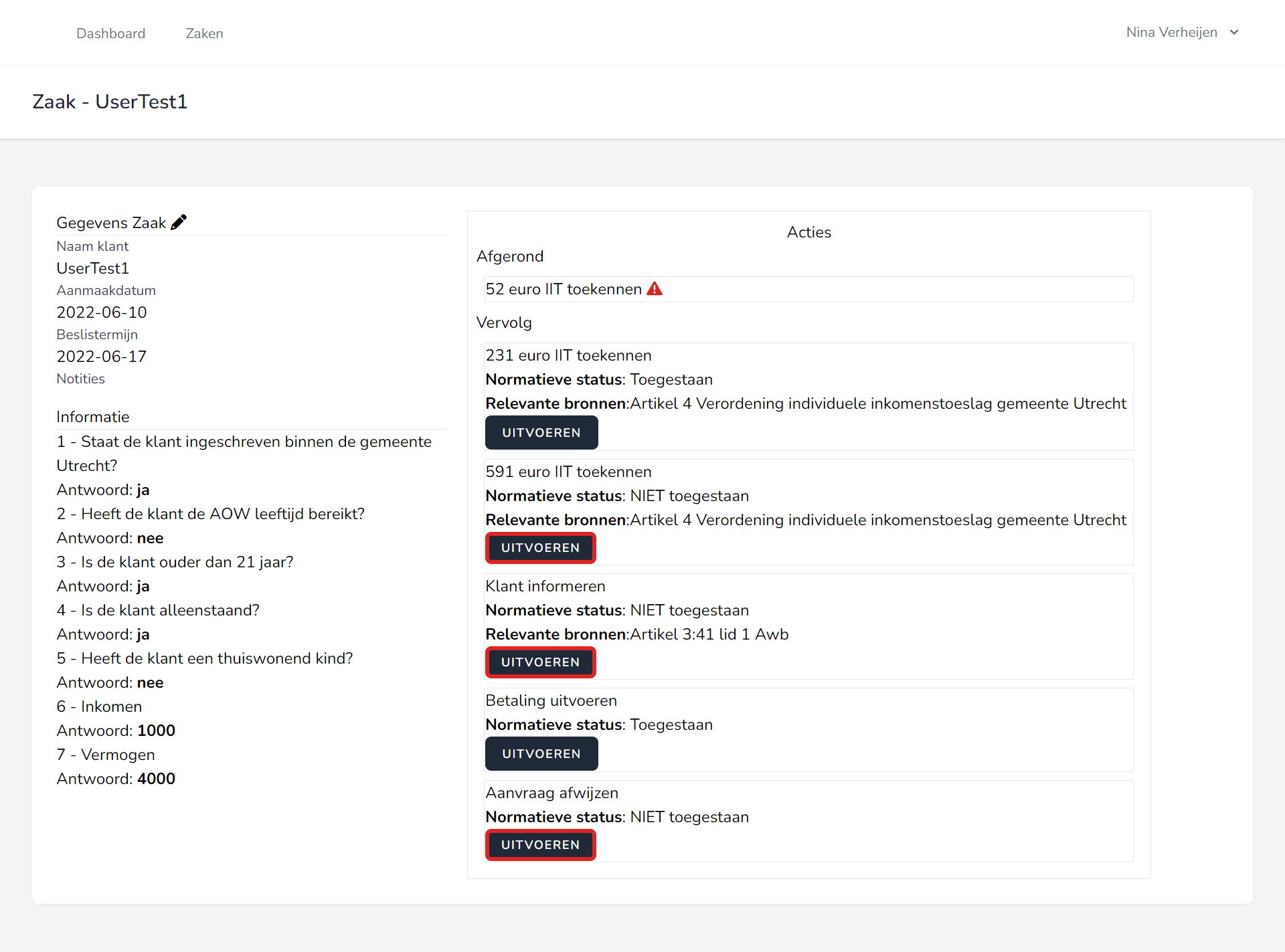}}
    \caption{Case overview page of the prototype.}
    \label{fig:case overview}
\end{figure}

\begin{figure}[htb!]
    \centering
    \setlength{\fboxsep}{0pt}
    \setlength{\fboxrule}{0.1pt}
    \fbox{\includegraphics[width=\textwidth]{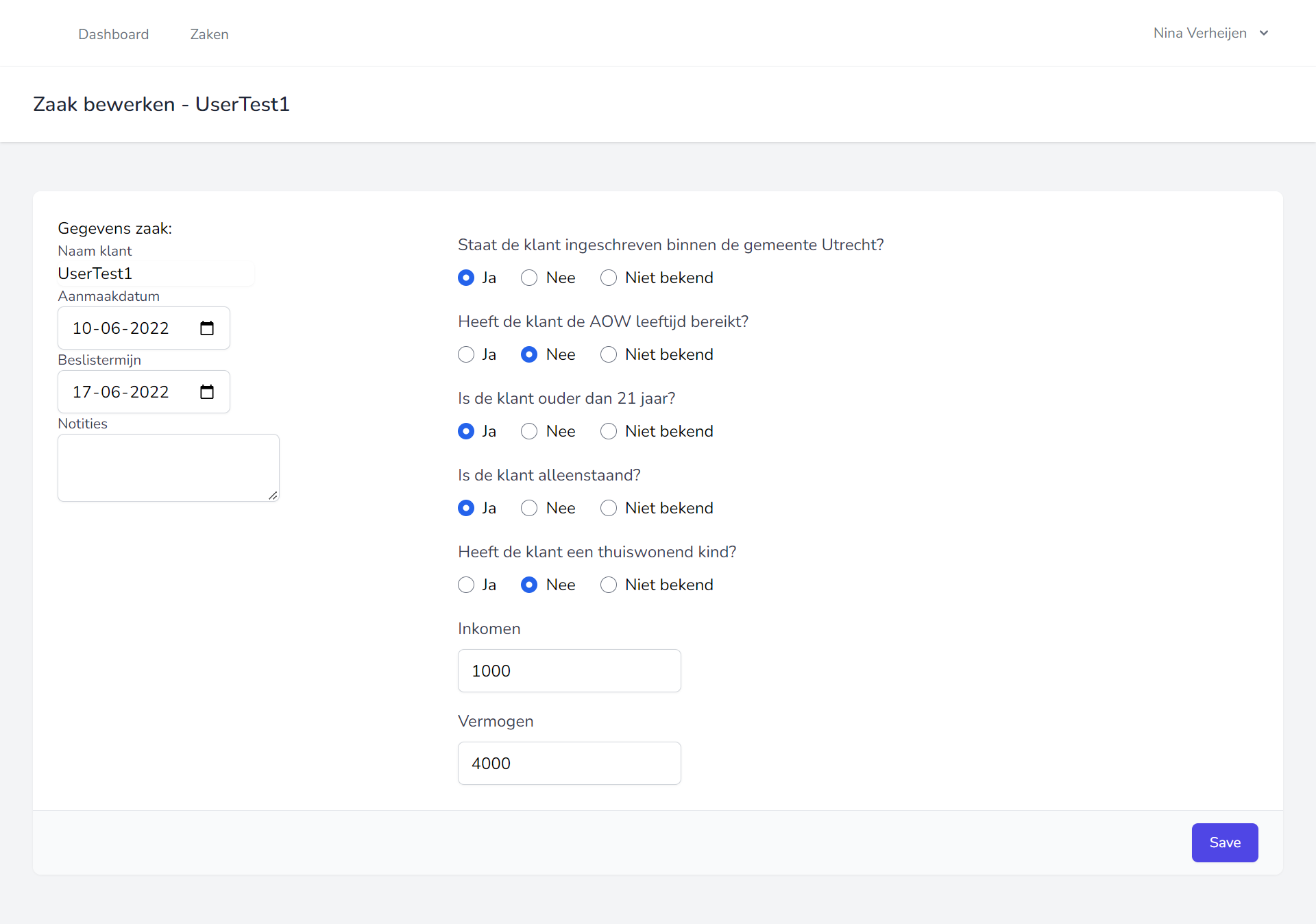}}
    \caption{Edit case page of the prototype.}
    \label{fig:Edit case}
\end{figure}

\begin{figure}[htb!]
    \centering
    \setlength{\fboxsep}{0pt}
    \setlength{\fboxrule}{0.1pt}
    \fbox{\includegraphics[width=\textwidth]{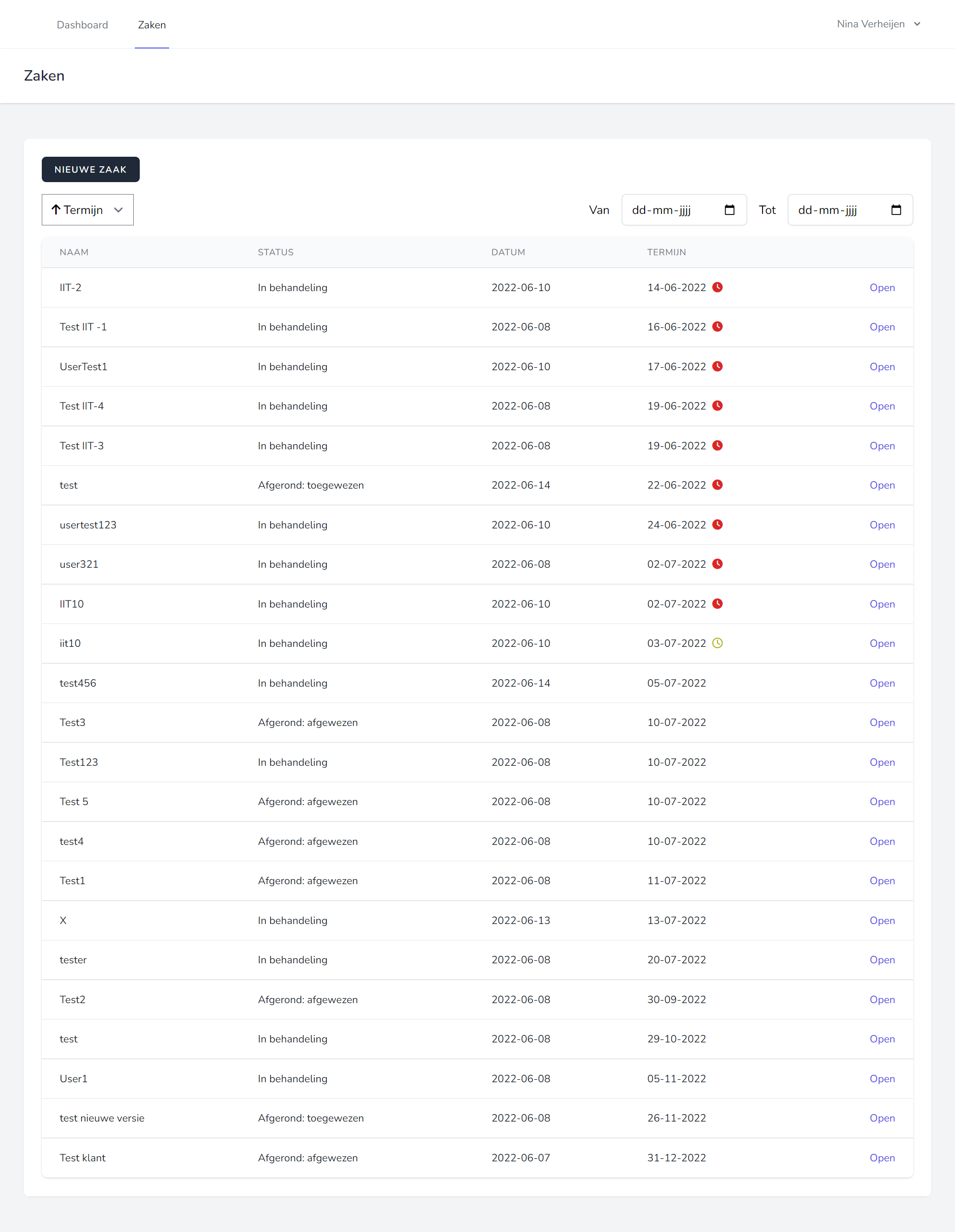}}
    \caption{Cases overview page of the prototype.}
    \label{fig:cases overview}
\end{figure}

\FloatBarrier


\end{appendices}

\end{document}